\documentclass{aa}
\usepackage{eht}
\usepackage[varg]{txfonts}
\usepackage{graphicx}
\usepackage[normalem]{ulem}
\usepackage{xcolor}
\usepackage{multirow}
\usepackage{multicol}
\usepackage{array}
\usepackage{comment}
\usepackage{bm}
\usepackage{tabularx}
\usepackage{booktabs}      %
\usepackage{pifont}         %

\usepackage[table]{xcolor} %
\definecolor{tred}{rgb}{1, 0.88, 0.88}
\definecolor{tgreen}{rgb}{0.88, 1, 0.88}
\definecolor{torange}{rgb}{1, 0.55, 0} 
\definecolor{tyellow}{rgb}{1, 1, 0.8}

\usepackage{hyperref}
\usepackage{float}
\hypersetup{
     colorlinks = true,
     linkcolor = {blue!50!black},
     anchorcolor = blue,
     citecolor = {blue!50!black},
     filecolor = blue,
     urlcolor = {blue!50!black}
}

\usepackage{orcidlink}
\let\orcid\orcidlink

\graphicspath{{figures}{.}}
\newcommand{\mycomment}[1]{}

\newcommand{\nonnegR}{\mathbb{R}_{\ge 0}}
\newcommand{\positiveR}{\mathbb{R}_{> 0}}
\newcommand{\natastron}{Nat. Astron.}

\begin{document}

\title{Optimal transport regularized dynamic radio interferometric reconstruction}

\subtitle{}

\author{
Andy Nilipour \orcid{0000-0002-5956-5167} \inst{\ref{cavendish}, \ref{kicc}, \ref{Yale}, \ref{naoj}} \and
Kotaro Moriyama \orcid{0000-0003-1364-3761} \inst{\ref{iaa-csic}, \ref{naoj_mizusawa}} \and
Shiro Ikeda \orcid{0000-0002-2462-1448} \inst{\ref{ism}, \ref{naoj}, \ref{sokendai_ism}, \ref{kavli_ipmu}} \and
Kazunori Akiyama \orcid{0000-0002-9475-4254} \inst{\ref{heriot-watt}, \ref{mithaystack}, \ref{naoj_mizusawa}, \ref{cfa}} \and
Mareki Honma \orcid{0000-0003-4058-9000} \inst{\ref{naoj_mizusawa}, \ref{sokendai_astro}, \ref{u-tokyo}}
}

\institute{Cavendish Astrophysics, University of Cambridge, Madingley Road, Cambridge CB3 0HA, UK, \label{cavendish} \email{andynilipour@gmail.com} \and Kavli Institute for Cosmology, University of Cambridge, Madingley Road, Cambridge CB3 0HA, UK \label{kicc} \and 
Department of Astronomy, Yale University, 219 Prospect Street, New Haven, CT 06511 \label{Yale} \and
National Astronomical Observatory of Japan, 2-21-1 Osawa, Mitaka, Tokyo 181-8588, Japan \label{naoj} \and
Instituto de Astrofísica de Andalucía-CSIC, Glorieta de la Astronomía s/n, E-18008 Granada, Spain \label{iaa-csic}  \and
Mizusawa VLBI Observatory, National Astronomical Observatory of Japan, 2-12 Hoshigaoka, Mizusawa, Oshu, Iwate 023-0861, Japan \label{naoj_mizusawa} \and
The Institute of Statistical Mathematics, 10-3 Midori-cho, Tachikawa, Tokyo, 190-8562, Japan \label{ism} \and
Statistical Science Program, The Graduate University for Advanced Studies (SOKENDAI), 10-3 Midori-cho, Tachikawa, Tokyo 190-8562, Japan \label{sokendai_ism} \and
Kavli Institute for the Physics and Mathematics of the Universe, The University of Tokyo, 5-1-5 Kashiwanoha, Kashiwa, 277-8583, Japan \label{kavli_ipmu} \and
Institute of Sensors, Signals and Systems, Heriot-Watt University, Edinburgh EH14 4AS, United Kingdom \label{heriot-watt} \and
Massachusetts Institute of Technology Haystack Observatory, 99 Millstone Rd, Westford, MA 01886, USA \label{mithaystack} \and
Center for Astrophysics $|$ Harvard \& Smithsonian, 60 Garden Street, Cambridge, MA 02138, USA \label{cfa} \and
Astronomical Science Program, The Graduate University for Advanced Studies (SOKENDAI), 2-21-1 Osawa, Mitaka, Tokyo 181-8588, Japan \label{sokendai_astro} \and
Department of Astronomy, Graduate School of Science, The University of Tokyo, 7-3-1 Hongo, Bunkyo-ku, Tokyo 113-0033, Japan \label{u-tokyo}
}

\authorrunning{
Nilipour et al.}
\titlerunning{Optimal transport regularized dynamic radio interferometric reconstruction}

\date{Submitted to A\&A on 9 July 2026; Received on 9 July 2026} %

\abstract{We propose a new technique for dynamic video reconstruction of radio interferometric data using the optimal transport (OT) distance as a regularizer within the regularized maximum likelihood (RML) framework. 
Compared to other commonly used regularization terms in RML and related prior-based approaches, the OT distance more naturally captures information including physical motion and promotes coherent dynamics.
This property makes the OT regularization particularly suitable for dynamic imaging of astrophysical phenomena on minute timescales. 
We perform a series of tests on simulated very long baseline interferometric (VLBI) observations of Sagittarius A$^\ast$ ({\sgra}), the supermassive black hole at the Galactic Center, with the Event Horizon Telescope (EHT), using both geometric and general relativistic magnetohydrodynamic models. 
We demonstrate that adding an OT regularization term yields significant improvements in dynamic reconstructions compared to those without it, as quantified by better recovery of orbital motion and higher dynamic cross-correlation metrics.
Continued developments in OT theory across various fields, proposed extensions to the EHT network, and broad applicability to other VLBI networks make it so that OT regularization for dynamic radio interferometric reconstructions will continue to be utilized and improved in the future.}


\maketitle
\nolinenumbers

\section{Introduction}\label{sec:intro}

Investigating horizon-scale emission near black holes is a prominent theme in contemporary astronomy and physics. One of the observational projects at the forefront of this pursuit is the Event Horizon Telescope (EHT), a global very long baseline interferometry (VLBI) array that links millimeter-wavelength radio telescopes to achieve a theoretical angular resolution of $\sim$25 \uas. Utilizing this capability, the EHT Collaboration produced the first horizon-scale images of the supermassive black hole at the center of the nearby radio galaxy M87 \citep{EHT_M87_PaperI, m87_2018, m87_variability}, and that at the center of the Milky Way, Sagittarius A$^\ast$ ({\sgra};\ \citealt{EHT_SgrA_PaperI}). Among all known black hole candidates, {\sgra} has the largest angular size of the black hole shadow (approximately 50 \uas;\ \citealt{EHT_SgrA_PaperIII, EHT_SgrA_PaperIV}). 
Its proximity allows for precise and accurate measurements of its mass and distance using the orbits of stars (e.g.,\ \citealt{Ghez2008, gravity18, gravity20}), enabling precise tests of general relativity \citep{EHT_SgrA_PaperVI}.

Because of its relatively low mass, {\sgra} is also known to exhibit pronounced variability on timescales ranging from minutes to hours. Multiwavelength observations have revealed flaring activity in infrared and X-ray bands (e.g.,\ \citealt{baganoff_rapid_2001, genzel_near-infrared_2003, marrone_x-ray_2008, yusef-zadeh_simultaneous_2009}), as well as millimeter-wavelength variability consistent with the timing of EHT observing campaigns \citep{wielgus2022, EHT_SgrA_PaperII}. Such rapid variability is thought to be closely linked to nonthermal emission processes near the event horizon and to dynamical structural changes within the accretion flow. Owing to these characteristics, {\sgra} provides one of the best laboratories for investigating the origin and physics of multiwavelength flares in the immediate vicinity of the event horizon. 

In VLBI, the Fourier components of the sky brightness distribution are measured only at discrete spatial frequencies $(u,v)$ corresponding to projected baseline vectors between pairs of antennas. To mitigate sparse sampling of the $(u,v)$ plane, VLBI imaging typically relies on Earth's rotation to continuously change the projected baselines between antennas, populating the $(u,v)$ plane with elliptical tracks as observations are taken over time. This technique assumes that the source image does not vary over the observation period. This assumption is violated for {\sgra}, posing significant challenges for static imaging and motivating the reconstruction of dynamic images, or videos (e.g.,\ \citealt{stewart_multiple-beam_2011, rau_radio_2012, bouman_reconstructing_2018, arras_unified_2019, arras_variable_2022, miller-jones_rapidly_2019, farah_selective_2022, muller_dog-hit_2022, muller_dynamic_2023, knollmuller_resolving_2023, roelofs2023, mus_new-generation_2024, mus2024, Foschi2025}). Such dynamic reconstructions of {\sgra} are expected to provide new observational constraints on the dynamics of accretion flows and magnetic fields near black holes and measurements of the spacetime around black holes (e.g.,\ \citealt{EHT_SgrA_PaperI, EHT_SgrA_PaperVII,eht2024_midrange}).

One straightforward approach to dynamic imaging is to extend the Regularized Maximum Likelihood (RML) methods of static imaging, which fit source image pixels directly to the data under constraints imposed by regularization terms (e.g.,\ \citealt{frieden1972, narayan1986, honma_super-resolution_2014, Chael2016}). These regularizers impose constraints on the characteristics of the spatial intensity distributions, such as smoothness and sparsity \citep{akiyama2017}. In the dynamic case, the observation data are divided into frames, resulting in both sparser $(u,v)$ coverage with limited hour-angles and shorter image integration time for each frame. Thus, it is necessary to include additional terms that constrain the temporal evolution across frames, such as continuity between frames \citep{johnson_dynamical_2017}. These additional dynamic regularization terms encourage the smooth temporal evolution of the intensity distributions throughout the observation. Existing dynamic regularizers typically rely on pixel-wise differences between consecutive frames or, in some cases, assume a specific flow field to describe the evolution of brightness. A related approach, the new-generation Maximum Entropy Method (ngMEM;\ \citealt{mus_new-generation_2024}), similarly utilizes a pixel-wise entropy term to enforce a minimum time variation compatible with the data.

In this article, we propose a new dynamic regularizer based on optimal transport (OT).
OT was first formulated by Gaspard Monge, who sought to move earth from one location to another with minimal physical work \citep{monge_memoire_1781}, and was later generalized by \citet{kantorovich_translocation_1941}. 
The OT problem defines a metric, the Wasserstein distance, as the minimum cost to transport mass (e.g.,\ resources or probability density) between two probability distributions (e.g.,\ \citealt{villani_optimal_2008, kolouri_optimal_2017}). The definition of ``probability distribution'' is flexible, and for the purposes of imaging, we define it as discrete, normalized, non-negative images.

The idea of OT is widely used in computer graphics and vision \citep{bonneel_survey_2023}, including as an objective function in machine learning \citep{frogner_learning_2015, arjovsky_wasserstein_2017}, as a method for image interpolation \citep{haker_optimal_2003, hug_multi-physics_2015}, as a constraint in inverse imaging \citep{metivier_graph_2019, lee_unbalanced_2020, gorszczyk_graph-space_2021}, and, most relevant to our work, as a regularizer for sparse multi-task regression \citep{janati_wasserstein_2019}. OT has also been utilized in astronomy, although to a lesser extent, with examples including the reconstruction of the cosmological primordial density field \citep{levy_fast_2021, nikakhtar_optimal_2023}, super-resolution imaging of noisy, sparse data \citep{rawson_optimal_2022}, and interpolation of supernova spectral time series \citep{ramirez_novel_2024}. 

Given the modern development of OT-based methods, it is natural to use a version of the Wasserstein distance as a regularizer for dynamic imaging of radio interferometric data. 
The Wasserstein distance goes beyond pixel-wise differences by accounting for the underlying image domain information, and thus can inherently leverage information about the physical motion between images, making it particularly useful in cases where coherent motion is expected. It also, unlike flow-based regularizers, does not make strong assumptions about the existence of stable velocity fields, which may not hold for physical systems such as {\sgra}. By incorporating efficient transport based on the OT problem, we aim to detect dynamics in the immediate vicinity of black holes.

This article presents a proof-of-concept demonstration of the OT regularizer for radio interferometry through a series of simulated observations based on the 2017 April 11 EHT $(u,v)$ coverage of {\sgra}. We demonstrate how the OT regularizer operates within the context of dynamic imaging, discuss the key properties that make it an effective dynamic regularizer, and show that our imaging pipeline enables recovery of coherent dynamical features. The implementation of the OT regularizer is provided as a new extension to the official EHT imaging software, \texttt{eht-imaging} (v1.2.4;\ \citealt{chael_interferometric_2018, chael_2025_14624987}), which we refer to as \texttt{ehtim+OT}.

The rest of the article is organized as follows. In Section \ref{sec:rml_and_ot}, we discuss in more detail the formulation of the radio interferometric imaging problem, RML methods, and the OT regularizer. In Section \ref{sec:motivation}, we explain the motivation behind using the OT regularizer for dynamic imaging. In Section \ref{sec:experiments}, we present reconstructions of simple simulated observations using the OT regularizer and compare them to reconstructions without the use of the OT regularizer, and we conclude in Section \ref{sec:conclusion}.
All validation tests establishing the reliability of \texttt{ehtim+OT} are described in detail in \citet{ehtim_official}.

\section{Dynamic Imaging with the Optimal Transport Regularizer}\label{sec:rml_and_ot}

\subsection{Radio Interferometric Measurement Equation for VLBI}\label{subsec:radio_interferometry}

After observing a celestial object with a network of radio antennas simultaneously (as, for example, part of an EHT imaging campaign), visibilities are computed from each pair of antennas. The ideal visibilities $V_{pq}$ measured by a baseline $\bm{b}_{pq}$ between stations $p$ and $q$ are related to the brightness distribution on the sky, $I(x,y)$, via the Fourier transform: \citep{thompson_interferometry_2017},
\begin{align}
    \label{eq:visibility}
    V_{pq} = \tilde{I}(u,v) = \int I(x,y) e^{-2\pi i (ux+vy)} dxdy.
\end{align}
Here, $(u,v)$ denotes the dimensionless coordinates of the projection of $\bm{b}_{pq}$ in the plane perpendicular to the line of sight, and $(x,y)$ is the angular coordinate in the sky. 

However, in practice, radio interferometric data are corrupted by a series of station-based and baseline-based propagation effects. These are naturally described in the framework of the Radio Interferometric Measurement Equation (RIME;\ \citealt{hamaker1996, Hamaker2000}). In this work, we restrict attention to total intensity only; under this assumption, the RIME reduces to the scalar relation \citep{smirnov2011, pesce2021},
\begin{align}
    V_{pq} = g_p g_q^* e^{i(\theta_p - \theta_q)} \tilde{I}(u,v) + n_{pq},
\end{align}
where $g_p e^{i\theta_p}$ is the complex gain for station $p$ in the chosen receptor channel, with amplitude $g_p$ and phase $\theta_p$, $\tilde{I}(u,v)$ are the ideal visibilities as in Equation \ref{eq:visibility}, and $n_{pq}$ is a complex noise term including thermal noise.

The data products we use in our imaging pipeline are the visibility amplitudes, $\lvert V_{pq} \rvert$, 
 and the closure products. While the visibility amplitudes remain sensitive to station amplitude gains, the closure quantities are robust to the gain terms \citep{chael_interferometric_2018}. The closure phases are formed by multiplying three complex visibilities in a closed triangle of baselines, which gives a visibility bispectrum,
 \begin{align}
     V_B = \lvert V_B\rvert e^{i\psi} = V_{12} V_{23} V_{31},
 \end{align}
 whose phase $\psi$ is independent of any of the individual station phase errors \citep{jennison1958, rogers1974, thompson_interferometry_2017}. The closure amplitudes are formed by taking ratios of visibility amplitudes in sets of four stations such that the individual station amplitude gains cancel out, as follows:
 \begin{align}
     \lvert V_C\rvert_a = \left\lvert \frac{V_{12} V_{34}}{V_{13}{V_{24}}} \right\rvert,\, \lvert V_C\rvert_b = \left\lvert \frac{V_{13} V_{24}}{V_{14}{V_{23}}} \right\rvert,\, \lvert V_C\rvert_c = \left\lvert \frac{V_{14} V_{23}}{V_{12}{V_{34}}} \right\rvert.\end{align}
Although the closure phases and amplitudes are insensitive to gain errors, they are still affected by thermal noise, and they contain less information about the source brightness distribution than the full visibilities.

\subsection{Regularized Maximum Likelihood Methods}\label{subsec:rml}

RML methods, which are now widely used in VLBI imaging, estimate the source image by minimizing an objective function,
\begin{equation}
\label{eq:staticRML}
\min_{\bm{I}} 
\Bigl[
\chi^2(\bm{I},\bm{V})
+ \sum_r \lambda_r R_r(\bm{I})
\Bigr],
\end{equation}
where the first term represents the error between the reconstructed image $\bm{I}$ and the data products $\bm{V}$, and the second term, which is the sum of regularizers $R_r$ weighted by hyperparameters $\lambda_r$, controls certain features in the image. In the static imaging case, common regularizers include total variation or total squared variation to control the smoothness of the image, the $L_1$ norm to control the sparsity, and the negative entropy term for the maximum entropy method (e.g.,\ \citealt{wernecke_maximum_1977, wiaux_compressed_2009, honma_super-resolution_2014, kuramochi_superresolution_2018}).

In the case of dynamic imaging, we associate a data error term and static regularizer terms for each reconstructed frame $\bm{I}_m$ of the video $\{\bm{I}_m\}$, where $m = 1,\ldots,M$ indexes the $M$ video frames.
The objective function becomes
\begin{equation}
\label{eq:dynamicRML}
\min_{{\{\bm{I}_m\}}} 
\Bigl[
\chi^2(\{\bm{I}_m\},\bm{V})
+
\sum_{r,m} \lambda_r R_r(\bm{I}_m) + 
\sum_s \lambda^{\rm dyn}_s R^{\rm dyn}_s(\{\bm{I}_m\})
\Bigr],
\end{equation}
where $R^{\rm dyn}_s$ are dynamic regularizer terms, weighted by hyperparameters $\lambda^{\rm dyn}_s$, that act on the video as a whole. These dynamic regularizers constrain similarities between adjacent frames, features in the mean image, stable motion, or other properties of the reconstructed video \citep{johnson_dynamical_2017}. In this work, we introduce the OT regularizer as an additional dynamic term. The new regularizer has been implemented as an extension to the existing \texttt{eht-imaging} library \citep{chael_2025_14624987}.

\subsection{Optimal Transport Regularizer}\label{subsec:OT_regularizer}

OT defines a distance between two probability distributions. In the context of imaging, these probability distributions are non-negative images defined on a discrete grid of pixels where the total sum is normalized to 1. In particular, for this work, they are consecutive frames in a dynamic video reconstruction.

Let us start by describing the OT problem. Consider two normalized images represented as vectors $\bm{I}_k$ and $\bm{I}_{\ell}$, so that $\bm{I}_k, \bm{I}_{\ell} \in \nonnegR^N$, where $\sum_i I_{k, i} = \sum_j I_{\ell, j} = 1$, $N$ denotes the total number of pixels, and $I_{k, i}\ (I_{\ell, j})$ denotes the $i^{\rm th}\ (j^{\rm th})$ pixel of $\bm{I}_k\ (\bm{I}_\ell)$. Assume the pixel intensity values of $\bm{I}_k$ represent the mass distributed across the image. The goal of the problem is to optimally transport the mass in each pixel to match the target mass distribution $\bm{I}_\ell$, while minimizing the total cost of the transport. Specifically, let $C_{ij} \ge 0$ represent the cost to transport a unit of mass from $I_{k, i}$ to $I_{\ell, j}$, and let $P_{ij} \ge 0$ denote the amount of mass to be transported between these pixels. The corresponding optimization problem is formulated as follows:
\begin{equation}
\label{eq:OT}
    \min_{P \in \nonnegR^{N \times N}} \sum_{i = 1}^N \sum_{j=1}^N C_{ij} P_{ij},\hspace{.5em}
    \mbox{subject to}\hspace{.5em}\sum_{j=1}^N P_{ij} = I_{k, i},\ \sum_{i=1}^N P_{ij} = I_{\ell, j}.
\end{equation}

The constraints ensure that the total sum of mass is conserved. 
In the present balanced OT formulation, flux is redistributed between adjacent normalized frames under exact mass conservation. Therefore, strongly appearing or disappearing localized features are not modeled as true sources or sinks of intensity, but are instead approximated through redistribution of flux. This approximation is appropriate when frame-to-frame total flux changes are modest, but more general unbalanced OT formulations may be preferable when intrinsic flux creation or disappearance is significant.

Let $P^\ast$ be the optimal matrix of the above problem. Then, the corresponding total cost is
\begin{equation}
    W(\bm{I}_k, \bm{I}_\ell) = \sum_{i = 1}^N \sum_{j=1}^N C_{ij} P^\ast_{ij}.
\end{equation}
In the case where the cost matrix $C$ defines a metric on $\mathbb{N}(N) = \{1, \hdots, N\}$, this defines the so-called Wasserstein metric \citep{villani_optimal_2008}. The matrix $C$ can be any metric, and we utilize a squared Euclidean distance between pixels on the two-dimensional image grid, although this can possibly be modified. 

The Wasserstein metric is calculated exactly with a linear programming optimizer, but the computation is expensive and impractical for large $N$. Instead, we use the Sinkhorn distance \citep{cuturi_sinkhorn_2013}, which regularizes the Wasserstein metric with an entropy term weighted with $\varepsilon>0$ as follows:
\begin{align}
    \label{eq:ot_entropy}
    &\min_{P \in \positiveR^{N \times N}} \left[\sum_{i = 1}^N \sum_{j=1}^N C_{ij} P_{ij} - \varepsilon\sum_{i = 1}^N \sum_{j=1}^N P_{ij} (1-{\log} P_{ij})\right],\\\nonumber
    &\mbox{subject to}\hspace{1em}\sum_{j=1}^N P_{ij} = I_{k, i},\ \sum_{i=1}^N P_{ij} = I_{\ell,j}.
\end{align}
Let $P^\varepsilon$ be the optimal matrix of the above problem. Then, the Sinkhorn distance is given by
\begin{equation}\label{eq:sinkhornDistance}
W_\varepsilon(\bm{I}_k, \bm{I}_\ell) = \sum_{i = 1}^N \sum_{j=1}^N C_{ij} P^\varepsilon_{ij} - \varepsilon\sum_{i = 1}^N \sum_{j=1}^N P^\varepsilon_{ij} (1-{\log} P^\varepsilon_{ij}). 
\end{equation}
While $W_\varepsilon(\bm{I}_k, \bm{I}_\ell)$ is always uniquely minimized and approaches $W(\bm{I}_k, \bm{I}_\ell)$ as $\varepsilon \to 0$, the entropically regularized optimal transportation matrix $P^{\varepsilon}$ is strictly positive and is generally less sparse than the corresponding $P^\ast$. Although the Sinkhorn distance does not satisfy the coincidence property $\bm{I}_k = \bm{I}_\ell \Leftrightarrow W_\varepsilon(\bm{I}_k, \bm{I}_\ell) = 0$, all other metric properties, namely positivity, symmetry, and the triangle inequality, are satisfied. The Sinkhorn distance can be easily converted into a proper metric, or even be generalized to non-normalized distributions (e.g.,\ \citealt{frogner_learning_2015, chizat_2017_sclaing}), but these corrections are unnecessary for our purpose of the dynamic regularizer. An efficient algorithm exists for calculating the Sinkhorn distance and its gradient \citep{luise_differential_2018}, making it suitable for RML optimization methods that require the gradient of the objective function. We define the OT regularizer as the average of the Sinkhorn distance between consecutive frames; that is,
\begin{equation}\label{eq:OTreg}
R_{\text{OT}}(\{\bm{I}_m\}) = \frac{1}{M-1}\sum_{m=1}^{M-1} W_\varepsilon(\bm{I}_m, \bm{I}_{m+1}).
\end{equation}

\begin{figure*}[htb]
    \centering
    \includegraphics[width=0.9\linewidth]{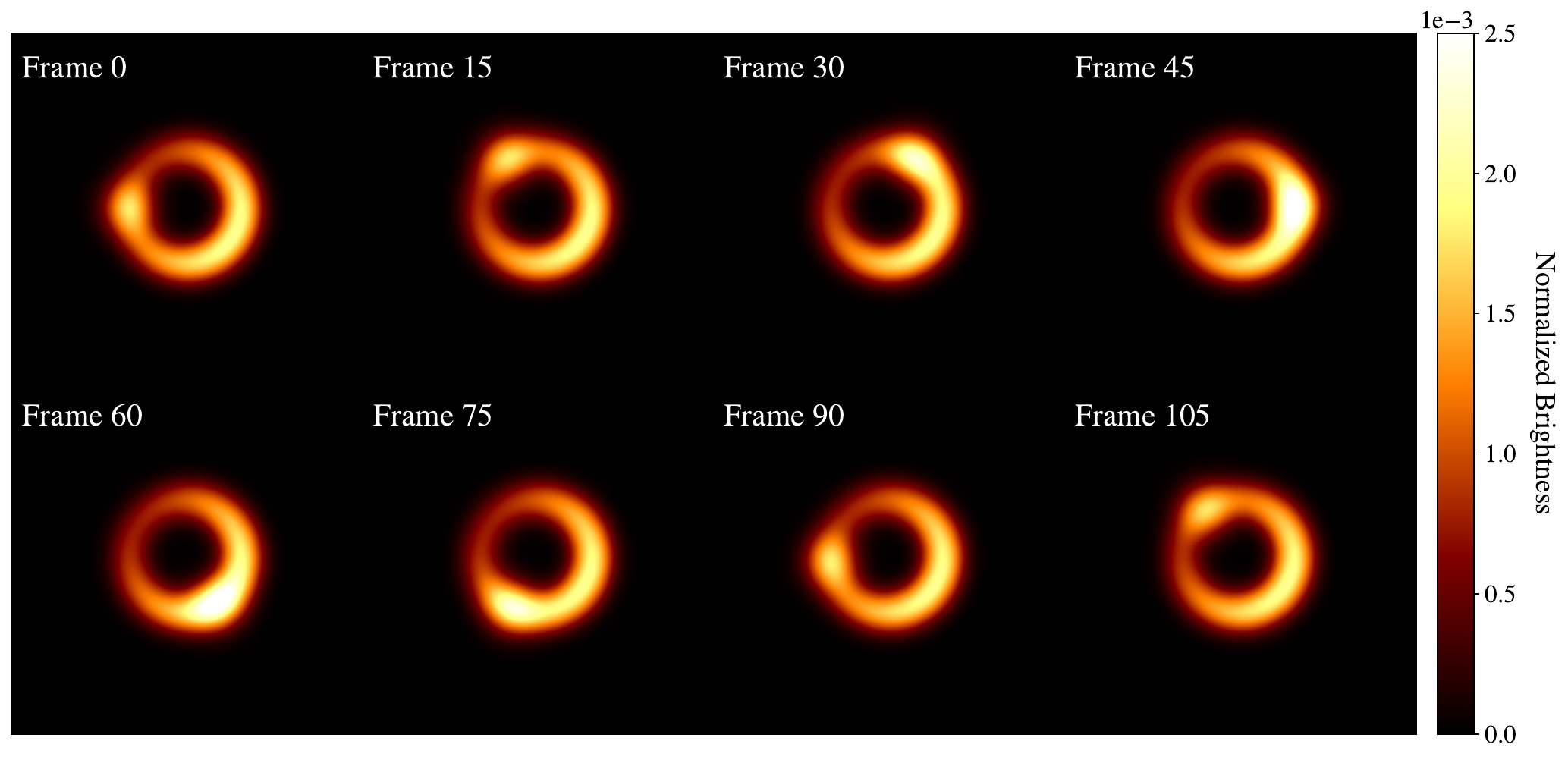}\\ 
    \includegraphics[width=0.9\linewidth]{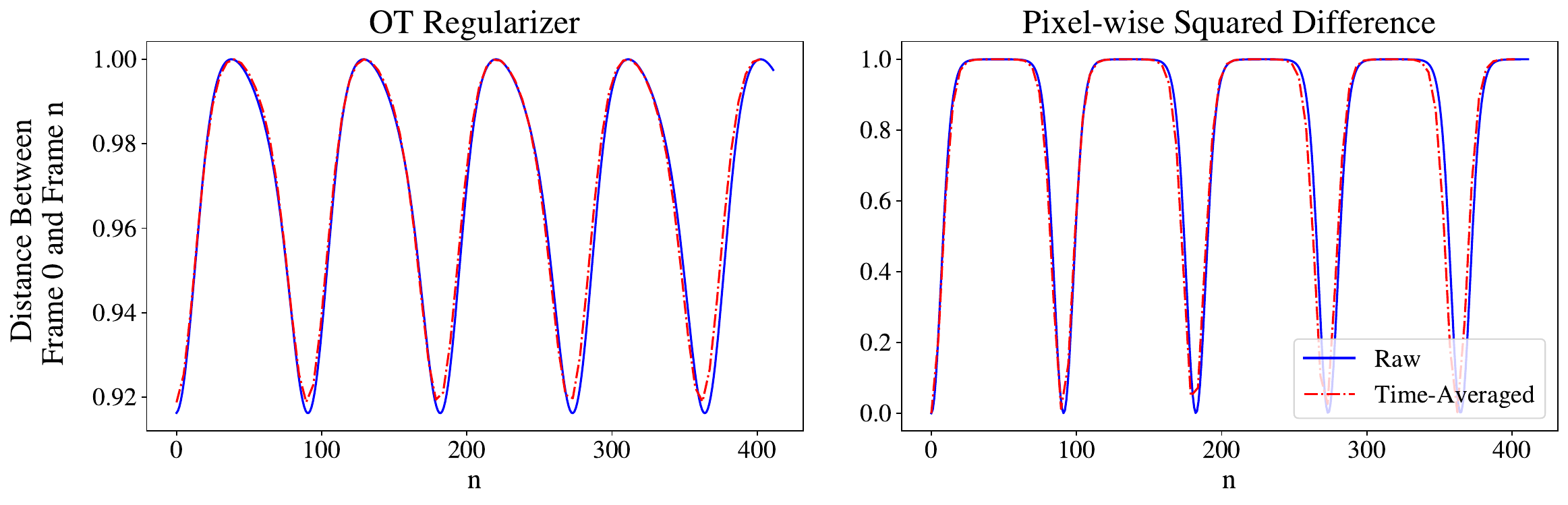}\\
    \caption{Comparison between the Sinkhorn and squared difference distances. The top panel shows frames of a simple geometric model consisting of a hotspot of constant brightness orbiting a static \texttt{m-ring} in the counterclockwise direction with a period of approximately 90 frames. The total brightness in each frame is normalized to unity. The bottom panel shows the behavior of the Sinkhorn distance (left) and the squared difference distance (right) for this model. The solid blue lines denote the distance between the $0^\text{th}$ and the $n^\text{th}$ frames. The broken red lines denote the same distances, but calculated for a video that has been time-averaged with a rolling mean of ten frames. The distances are normalized to a maximum of unity.}
    \label{fig:otpwHotspot}
\end{figure*}

\begin{figure*}[htb]
    \centering
    \includegraphics[width=0.85\linewidth]{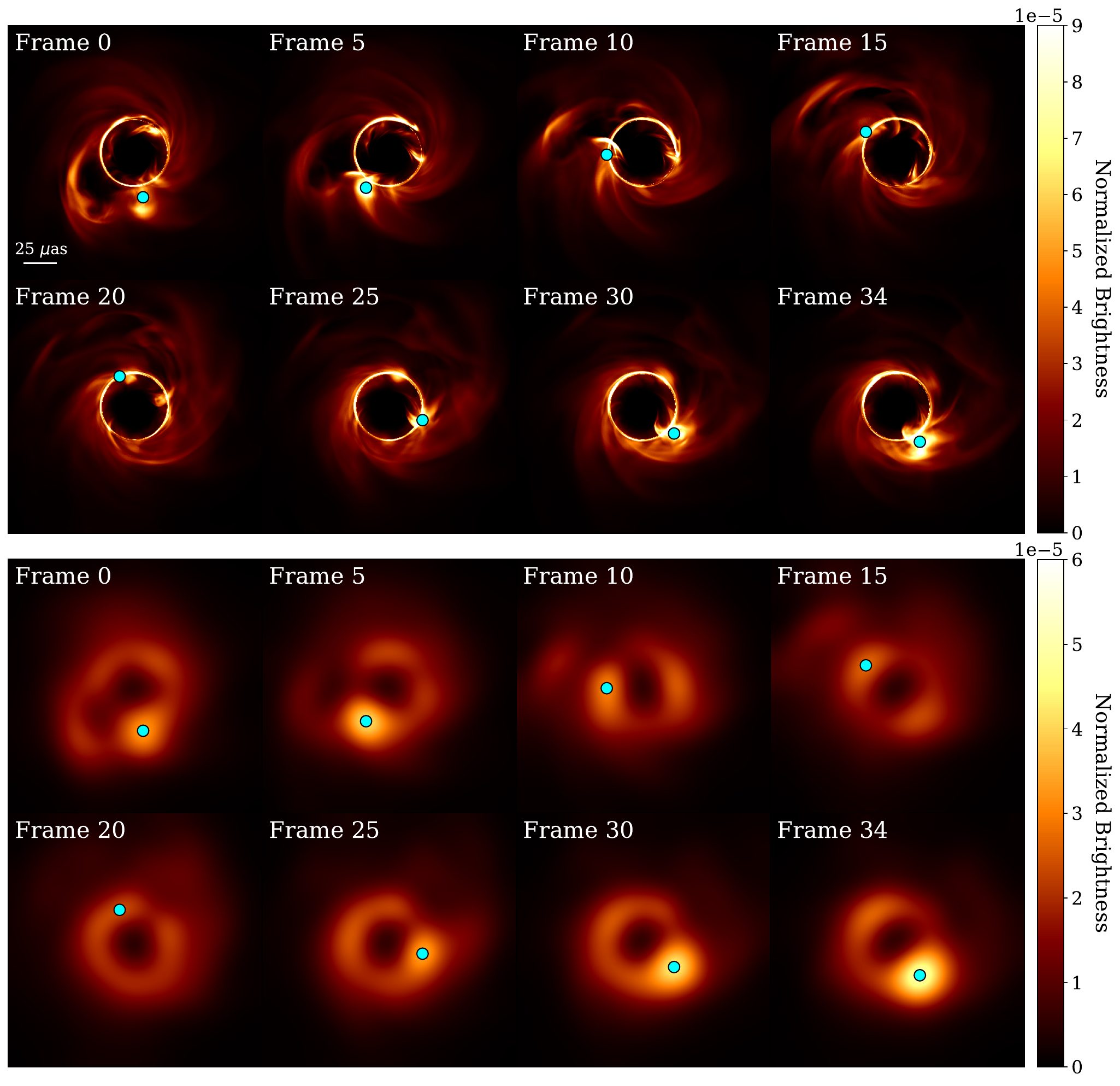}\\ 
    \includegraphics[width=0.8\linewidth]{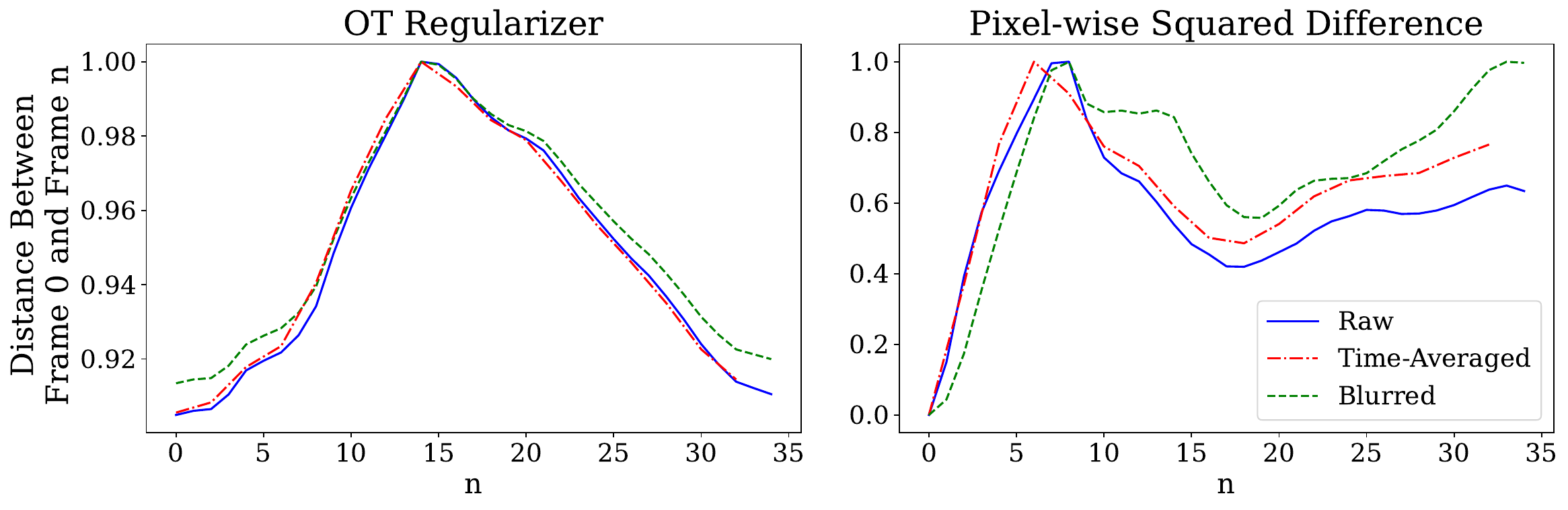}
    \caption{Same as Figure \ref{fig:otpwHotspot}, but for a rotating GRMHD model. We also show the GRMHD model blurred by a Gaussian kernel with a full width at half maximum (FWHM) of 25~\uas, which is approximately the nominal EHT resolution. The blue dot denotes the location of peak brightness of the blurred model. In the bottom panels, the Sinkhorn and squared difference distances are also calculated for the blurred model. The time-averaging for this model is performed with a rolling mean of two frames.}
    \label{fig:otpwGRMHD}
\end{figure*}

\section{Motivation for the OT Regularizer}\label{sec:motivation}

\subsection{Comparison to Pixel-Wise Distances}

The dynamic imaging method of \cite{johnson_dynamical_2017} is based on the same framework as Equation~\ref{eq:dynamicRML}, but they use different dynamic regularizers different than the OT regularizer. The first regularizer they propose is the distance regularizer, defined as
\begin{equation}
    R_{\Delta t}(\{\bm{I}_m\}) = \frac{1}{M-1}\sum_{m=1}^{M-1} \mathcal{D}(\bm{I}_m, \bm{I}_{m+1}),
\end{equation}
where the distance function $\mathcal{D}$ is given either by the total squared difference,
\begin{equation}\label{eq:pixelwiseSq}
\mathcal{D}_\text{sq}(\bm{I}_m,\bm{I}_{m+1}) = \sum_{i=1}^N (I_{m, i} - I_{m+1,i})^2,
\end{equation}
or by the relative entropy,
\begin{equation}\label{eq:pixelwiseRelE}
\mathcal{D}_\text{KL}(\bm{I}_m,\bm{I}_{m+1}) = \sum_{i=1}^N I_{m, i} \ln\left ( \frac{I_{m, i}}{I_{m+1,i}}\right ).
\end{equation}
Both are defined as the sum of pixel-wise functions, and therefore, the encoded information about the underlying flux motion between images is limited. The second regularizer they propose is the sum of the distances between the individual frames and the average frame,
\begin{equation}
    R_{\Delta I}(\{\bm{I}_m\}) = \frac{1}{M}\sum_{m=1}^{M} \mathcal{D}(\bm{I}_{\rm avg}, \bm{I}_{m}),
\end{equation}
where $\mathcal{D}$ is taken to be the same pixel-wise distance used for $R_{\Delta t}$. These two regularizers involve one hyperparameter each, namely the weights $\lambda_{\Delta t}$ and $\lambda_{\Delta I}$. In some cases, however, it may be desirable to blur the images before computing $\mathcal{D}(\bm{I}_{m}, \bm{I}_{m+1})$ and $\mathcal{D}(\bm{I}_{\rm avg}, \bm{I}_{m})$. In such cases, the standard deviations of the circular Gaussian blurring kernels, $\alpha_{\Delta t}$ and $\alpha_{\Delta I}$, can also be included as additional hyperparameters. In our imaging experiments presented in Section \ref{sec:experiments}, we refer to $R_{\Delta t}$ and $R_{\Delta I}$ as the \textit{standard dynamic regularizers}.

The OT regularizer is closely related to the distance regularizer, as both are defined as the sum of distances between consecutive frames. However, we expect the OT regularizer to capture spatially correlated flow more effectively because the Sinkhorn distance reflects the global brightness distributions, while the distance regularizer compares images locally.

To demonstrate that the OT regularizer preserves more information about the frame-to-frame motion, we compare the behavior of the Sinkhorn distance in Equation \ref{eq:sinkhornDistance} and the squared difference distance in Equation \ref{eq:pixelwiseSq} on two models. The first model, shown in the top panel of Figure \ref{fig:otpwHotspot}, consists of a simple Gaussian component (a \textit{hotspot}) orbiting a static first-order \texttt{m-ring}, which is a ring of non-uniform brightness in azimuthal directions given by a first-order Fourier series \citep{johnson2020}. The second model, shown in the top panel of Figure \ref{fig:otpwGRMHD}, consists of a rotating general relativistic magnetohydrodynamic (GRMHD) model drawn from the ``Illinois v3'' library \citep{dhruv2025}.  The model is based on simulations run with the \texttt{kharma} code \citep{prather2024} and imaged with \texttt{ipole} \citep{moscibrodzka2018} using the \texttt{PATOKA} pipeline \citep{wong2022}.

The Sinkhorn and squared difference distances between the $0^{\rm th}$ and the $n^{\rm th}$ frames for the hotspot model are shown in the bottom panel of Figure \ref{fig:otpwHotspot}. Ideally, the distance value should reach its maximum at half the rotation period, when the hotspot is the furthest away from its original location, and its minimum at a full rotation period, when the hotspot has returned to its original location. Indeed, the distance between the hotspot and its original location is sinusoidal with respect to time, which is accurately reflected by the Sinkhorn distance in the bottom left panel. On the other hand, the squared difference distance loses information about the location of the hotspot when it moves away from the original position. At any frame for which there is no overlap between the extent of the hotspot in that frame and in the $0^{\rm th}$ frame, the squared difference distance will be maximal; thus, the sinusoidal shape is flattened near its peaks (at each half-period) in the bottom right panel. For this model, both the Sinkhorn and the squared difference distances appear to be largely unaffected by time averaging. We note that while the Sinkhorn distance appears to exhibit a smaller dynamic range ($\sim$8\% variation), this does not impact the optimization, because the absolute scaling of the regularizer and its gradient is controlled by the hyperparameter $\lambda_{\rm OT}$, which we survey over several orders of magnitude.

Figure \ref{fig:otpwGRMHD} shows similar behavior for one period of the GRMHD model. The Sinkhorn distance again peaks at approximately half the period and returns close to its initial value after a full period. On the other hand, the squared difference distance increases linearly for the first few frames but subsequently appears to lose information about the underlying physical motion of the model. We see behavior similar to that of the hotspot model when the GRMHD model is time-averaged, although the squared difference distance may be slightly less robust to time averaging beyond the first few frames. 

To show that the OT regularizer will not be affected by angular resolution, we also consider the GRMHD model blurred with a Gaussian kernel to the nominal EHT resolution of 25~\uas. We find that the behavior of the Sinkhorn distance for the blurred model is nearly identical to the unblurred model, indicating that it is independent of resolution. Meanwhile, the squared difference distance increases linearly for approximately the first ten frames, as it does for the unblurred model, but beyond the tenth frame, the blurred and unblurred distances do not align.

\subsection{Transportation Matrix Coherence}
\label{subsec:transportationMatrix}

\begin{figure}[htb]
    \centering
    \begin{minipage}[c]{0.9\linewidth}
        \includegraphics[width=\linewidth]{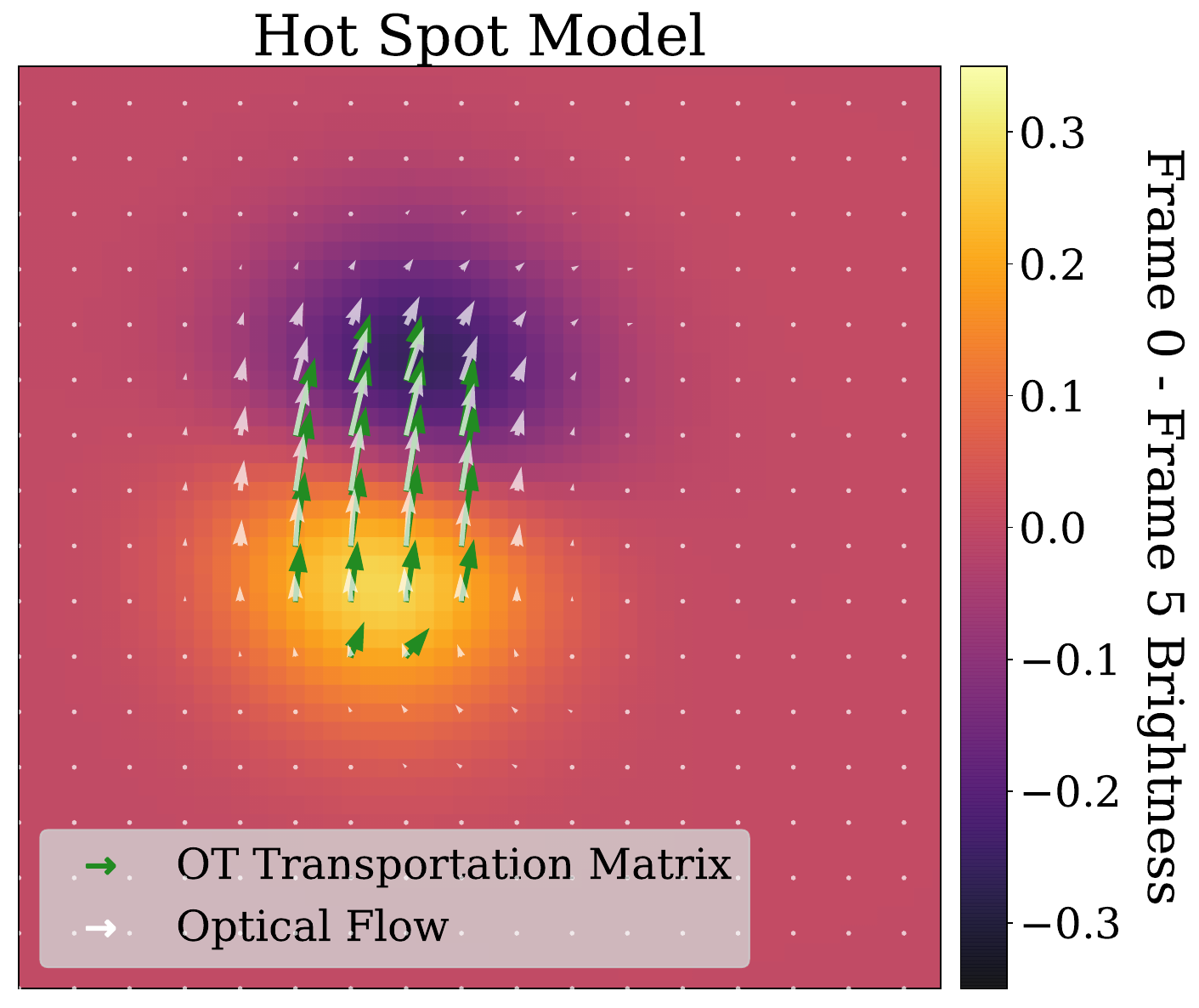}
    \end{minipage}\\
    \begin{minipage}[c]{0.9\linewidth}
        \includegraphics[width=\linewidth]{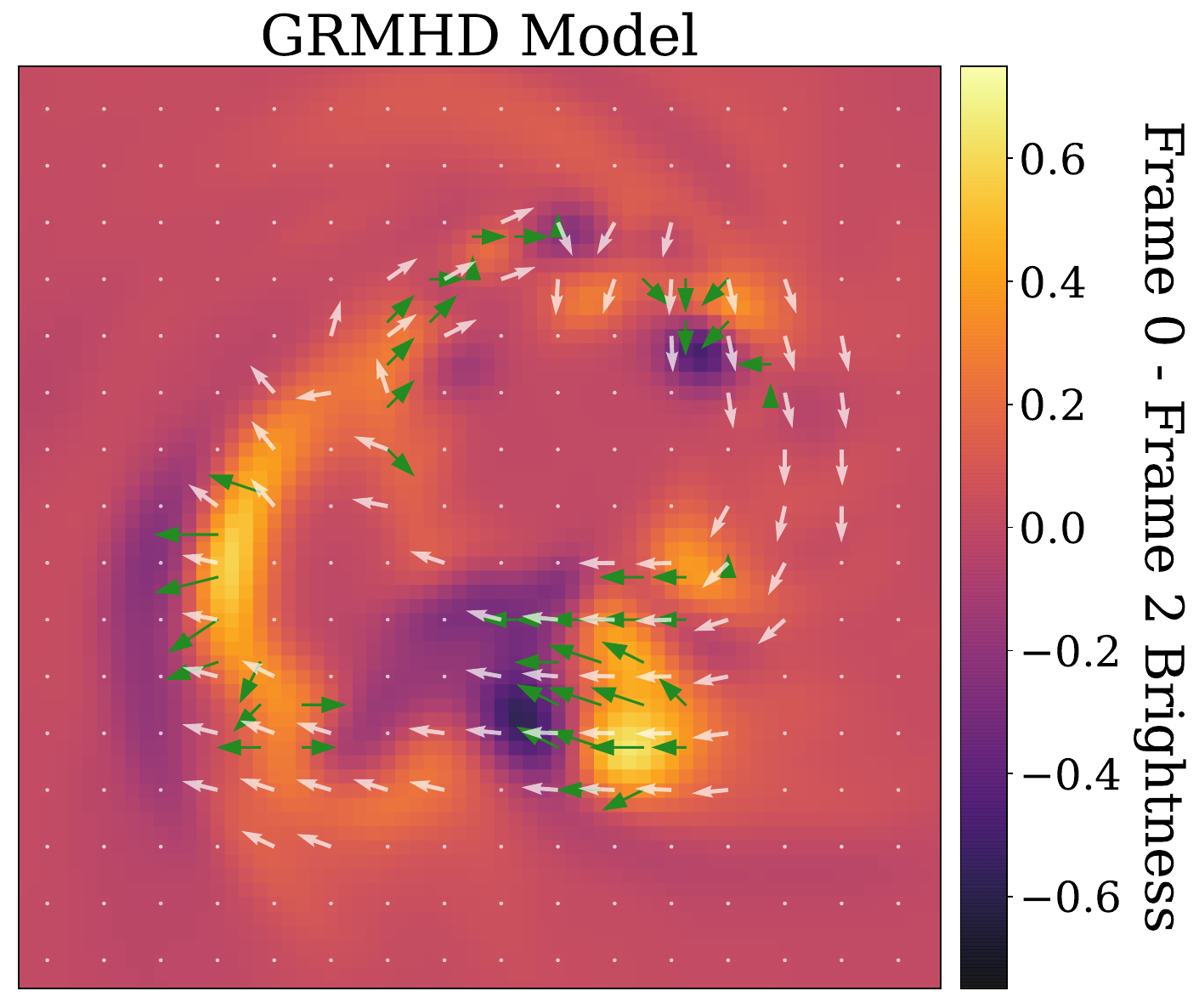}
    \end{minipage}
    \caption{Comparison between the optical flow field and the optimal transport matrix for two models. The background image of the top panel shows the difference between the normalized brightness of the $0^{\rm th}$ and the $5^{\rm th}$ frames of the same hotspot model as in Figure \ref{fig:otpwHotspot}, but zoomed in to focus on the hotspot location. Positive values indicate where the $0^{\rm th}$ frame is brighter and negative values indicate where the $5^{\rm th}$ frame is brighter, so that flux is moving from positive to negative values. The green arrows correspond to the transportation matrix of the Sinkhorn distance and point from pixels in the $0^{\rm th}$ frame towards the ``centers of mass'' of the flux transported from those pixels to the $5^{\rm th}$ frame. Arrows are spaced three pixels apart and are shown only for pixels for which the flux in the $0^{\rm th}$ frame is greater than 6.5 times the mean flux. The white arrows correspond to the optical flow field and are also spaced three pixels apart. The bottom panel shows the same information for the $0^{\rm th}$ and the $2^{\rm nd}$ frames of the same GRMHD model as in Figure \ref{fig:otpwGRMHD}, regridded and blurred to the nominal EHT resolution. We have also zoomed in on the central features. The green transportation matrix arrows are shown only on pixels for which the flux in the $0^{\rm th}$ frame is greater than 4.5 times the mean flux, and the white optical flow arrows are spaced four pixels apart. In the bottom panel, the white optical flow arrows are normalized in magnitude and are also shown only for pixels for which the flux in the $0^{\rm th}$ frame is greater than 2.5 times the mean flux.}
    \label{fig:arrowModel}
\end{figure}

The advantage of the OT regularizer lies in the motion information it inherently encodes between frames. This information is contained in the Sinkhorn distance transportation matrix, $P^{\varepsilon}$, which specifies the amount of mass to transport between pixels to optimally map one frame to the next. This nature is in some ways similar to that of the flow regularizer proposed by \citet{johnson_dynamical_2017}. The flow regularizer assumes that the evolving video has a stable flow vector field such that a given frame can be used to predict a later frame, and it penalizes the difference between the predicted and actual subsequent frames. The optical flow, which describes the apparent motion of brightness between two images or a series of images \citep{horn_determining_1981}, is in particular one way of estimating the flow vector field. OT theory has been proposed as an approach to better estimate the optical flow (e.g.,\ \citealt{haker_optimal_2003, kolesov_fire_2010}), thus motivating a comparison between the two.

In Figure \ref{fig:arrowModel}, we visualize the transportation matrix for the Sinkhorn distance between frames of a geometric and a GRMHD model as a vector field of mass displacements. The geometric model consists of a hotspot orbiting a ring of constant brightness, and the GRMHD model is that same as that of Figure \ref{fig:otpwGRMHD}. For each pixel, we compute the average direction of the mass from the first frame to the next frame; in other words, we find the center of mass of the transportation matrix row that corresponds to the pixel in the initial frame. We compare this with the optical flow derived with the algorithm introduced by \citet{farneback_two-frame_2003}, which uses polynomial expansions to estimate displacement fields between two frames at all points. 

The figure shows that the overall behaviors of the transportation matrix and the optical flow field are consistent, particularly for the hotspot model. For the GRMHD model, the optical flow field clearly exhibits rotational motion, and the transportation matrix also appears to capture some of this rotation. Although we do not expect the transportation matrix to match the optical flow field exactly, their approximate alignment is encouraging, since the optical flow field represents the velocity field of motion between images, which is the information that the standard dynamic regularizers, $R_{\Delta t}$ and $R_{\Delta I}$, do not encode. Another observation from the figure is that the transportation matrix tends to favor more local transport than the optical flow field. 

We expect that the accretion flow around a supermassive black hole, and most astrophysical phenomena for which dynamic interferometric reconstructions are relevant, may not be described by a simple velocity field, and that OT may capture the motion more effectively than the optical flow. The OT regularizer can be interpreted as a generalized version of the flow regularizer for which the flow is not assumed to be stable over time. Since the flow regularizer has not been utilized in works beyond \citet{johnson_dynamical_2017}, we focus on the more widely adopted pixel-wise standard regularizers, $R_{\Delta t}$ and $R_{\Delta I}$, for comparison with the OT regularizer on imaging tests in Section \ref{sec:experiments}.

\section{Imaging Demonstration}\label{sec:experiments}

In this section, we describe our imaging pipeline and perform imaging on a set of synthetic observations.

\subsection{Data}
\label{subsec:imaging_data}
We use synthetic observations that mimic the EHT observations on 2017 April 11, between 10.89 and 14.04\,UT.%
The data use a simple geometric model consisting of a hotspot orbiting a static first-order \texttt{m-ring}. The hotspot orbits in the clockwise direction with a period of 80 minutes. For the full imaging test in Section \ref{subsec:full_imaging}, we also validate our pipeline on a more physically motivated model consisting of a GRMHD model taken from the ``Illinois v3'' library \citep{dhruv2025}, with an added clockwise hotspot identical to that in the geometric model. These models correspond to the \texttt{mring+hsCW} and \texttt{grmhd2+hs} models of \citet{Dahale_2026}; see their Sections 2 and 4 for a complete description of how the data were generated. For all imaging demonstrations in this work, the data products we use are the visibility amplitudes, closure phases, and log closure amplitudes, as described in Section \ref{subsec:radio_interferometry}. In \citet{ehtim_official}, validation imaging tests are performed on the full standardized suite of synthetic data described in \citet{Dahale_2026}, including static geometric models, dynamic geometric models with coherent and incoherent motion, and additional GRMHD models.

\subsection{OT Regularizer Implementation}
\label{subsec:imaging_pipeline}

To image the data, we use the Python software \texttt{eht-imaging} \citep{chael_2025_14624987}. The standard dynamic regularizers are implemented using the existing \texttt{eht-imaging} implementation of \citet{johnson_dynamical_2017}, whereas the OT regularizer is implemented separately using the new method developed in this work.

We solve for the optimal solution of Equation \ref{eq:dynamicRML} by iteratively minimizing the objective function. We use the Limited-Memory BFGS (L-BFGS) algorithm \citep{byrd1995_lbfgs}, implemented in the optimization package \texttt{scipy.optimize.minimize} \citep{virtanen2020_scipy}. The optimization procedure requires evaluations of each term of Equation \ref{eq:dynamicRML} and their gradients. The gradients for the data terms are provided by \citet{chael_interferometric_2018} and the gradients for the standard dynamic regularizers are provided by \citet{johnson_dynamical_2017}. For the OT regularizer, we adapt the Sinkhorn algorithm \citep{cuturi_sinkhorn_2013, frogner_learning_2015} to calculate the Sinkhorn distance between each pair of adjacent frames. This algorithm casts Equation \ref{eq:ot_entropy} in its Lagrangian dual form and solves iteratively for the dual variables. These variables yield both the transportation matrix $P^\varepsilon_{ij}$, from which $W_\varepsilon(\bm{I}_m, \bm{I}_{m+1})$ can be calculated via Equation \ref{eq:sinkhornDistance}, and the gradient of the Sinkhorn distance with respect to either frame ($\partial W_\varepsilon/\partial \bm{I}_m$ or $\partial W_\varepsilon/\partial \bm{I}_{m+1}$).

RML imaging typically requires an exploration of thousands of imaging parameters, and determining the optimal set of parameters requires a metric of reconstruction fidelity when the ground truth video is known. We use the dynamic normalized cross-correlation, $\rho_{\rm DNX}$, which is included in the larger radio interferometric video evaluation framework developed by \citet{Dahale_2026}. This metric emphasizes fidelity of the \textit{dynamic component} of the reconstruction, which refers to the residuals from the pixel-wise median frame (the \textit{static component}). $\rho_{\rm DNX}$ is based on the normalized cross-correlation metric, which measures the similarity between two images as
\begin{equation}
    \rho(\bm{I}_k, \bm{I}_\ell) = \frac{\sum_{i} \left(I_{k, i} - \overline{\bm{I}_k}\right) \left(I_{\ell,i} - \overline{\bm{I}_\ell}\right)}{\sqrt{\sum_i \left(I_{k,i} - \overline{\bm{I}_k}\right)^2 \sum_i \left(I_{\ell,i} - \overline{\bm{I}_\ell}\right)^2}}, 
\end{equation}
where $\overline{\bm{I}_k}$ and $\overline{\bm{I}_\ell}$ are the mean values of $\bm{I}_k$ and $\bm{I}_\ell$.
 The dynamic normalized cross-correlation encodes the percentage of frames for which the normalized cross-correlation between the dynamic components of the reconstructed video and the ground truth video is above a frame-dependent threshold level. The threshold level depends on the instantaneous $(u, v)$ coverage of each frame. For exact definitions of $\rho_{\rm DNX}$ and the threshold values, we refer to \citet{Dahale_2026}. We use the dynamic normalized cross-correlation as the preferred evaluation metric because it accounts for the variable $(u,v)$ coverage of the EHT network and is sensitive to the dynamic component of the reconstruction, which is weak compared to the static component. While the precise value of $\rho_{\rm DNX}$ is not critical, we select the imaging parameters that maximize performance relative to this specific evaluation metric; the usefulness of this choice is further demonstrated by the hotspot detection presented in this paper and more comprehensively in \citet{ehtim_official}. The reconstruction that achieves the highest $\rho_{\rm DNX}$ value is referred to as the \textit{fiducial reconstruction}.

As another metric, we also evaluate how well the dynamic component of the reconstruction matches with that of the ground truth video. To do so, we extract the flux and position angle of the hotspot. We blur the dynamic component with a Gaussian kernel with a FWHM equal to that of the ground truth hotspot (20 $\mu$as). From the blurred frame, we mask out a circular filter of diameter 20 $\mu$as centered on the brightest pixel. We find that this size circular filter strikes a good balance between including the extent of the hotspot and excluding artifacts introduced by the imaging process that may inflate the reconstructed hotspot flux. The sum of the masked image is taken to be the hotspot flux. For the hotspot position angle, we calculate the center of mass of the masked image and take the angle from the center of the image. When applying this procedure to the ground truth video, the center of mass is generally equal to the brightest pixel, as expected for a purely Gaussian hotspot.

\subsection{Simple Imaging Test}
\label{subsec:simple_imaging}

\begin{figure*}[hbt]
    \centering
    \begin{minipage}[c]{1.0\linewidth}
        \includegraphics[width=\linewidth]{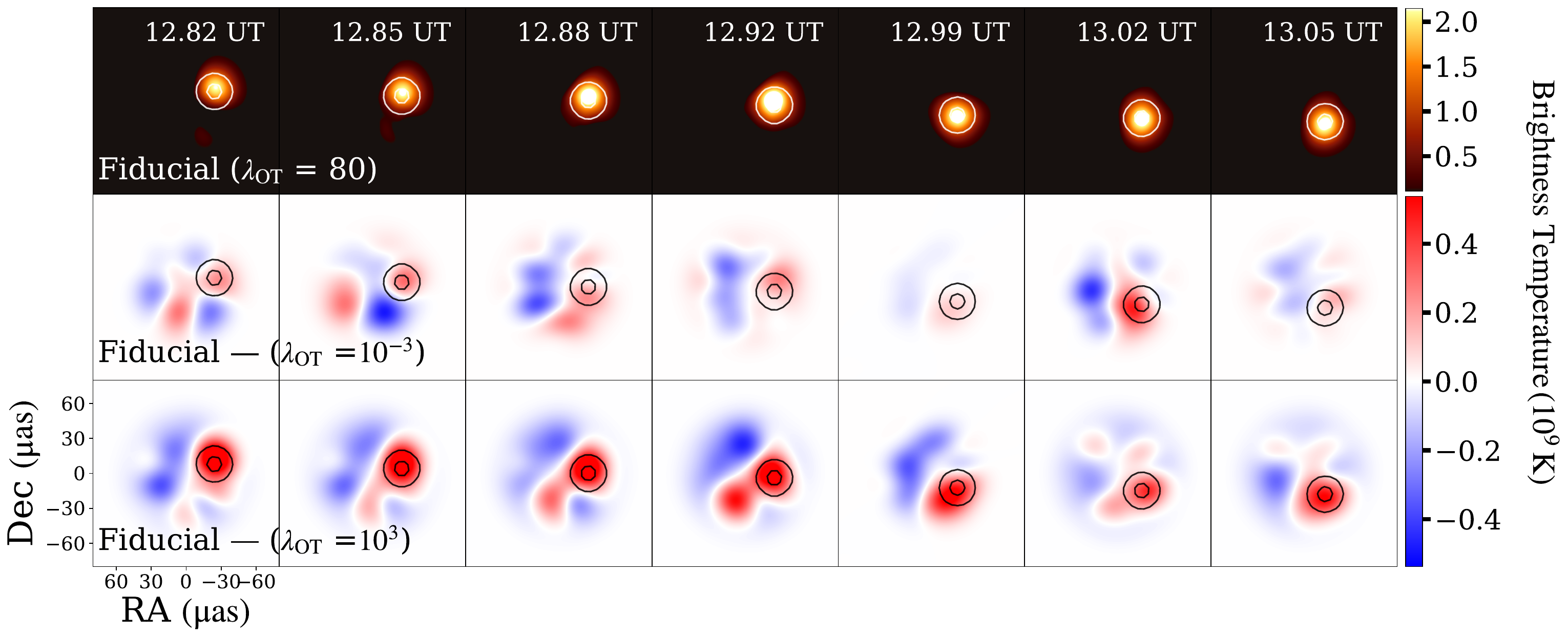}
    \end{minipage}
    \caption{Comparison of the fiducial reconstruction, with $\lambda_{\rm OT} = 80$, to reconstructions with a much lower and a much higher hyperparameter value for the simple imaging test. The top row shows the dynamic component of the fiducial reconstruction, blurred to the nominal EHT resolution, for a selected range of times. The middle and bottom panels show the difference between the blurred fiducial reconstruction and the blurred reconstructions with $\lambda_{\rm OT} = 10^{-3}$ and $\lambda_{\rm OT} = 10^3$, respectively. The positive (red) areas indicate where the fiducial reconstruction is brighter, and the negative (blue) areas indicate where it is less bright. The columns show consecutive frames, but not necessarily with equal time intervals, because there are time gaps in the data. The nominal time interval for each frame is two minutes (0.033 hours). The contours (white in the top row and black in the middle and bottom rows) denote the position of the ground truth hotspot.}
    \label{fig:test_recon}
\end{figure*}

\begin{figure*}[hbt]
    \centering
    \begin{minipage}[c]{1.0\linewidth}
        \includegraphics[width=\linewidth]{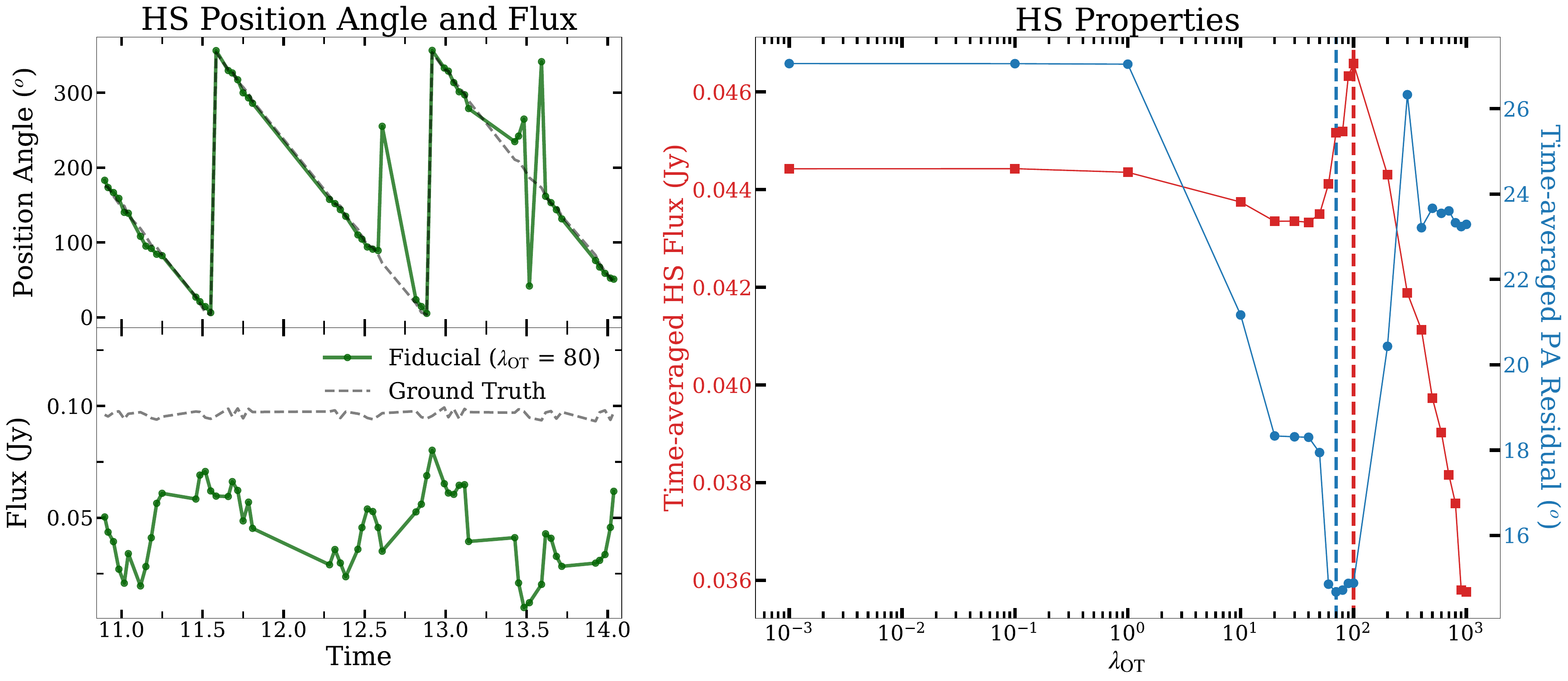}
    \end{minipage}
    \caption{Reconstruction hotspot properties for the simple imaging test. The top and bottom left panels compare the hotspot position angle and hotspot flux, respectively, of the fiducial reconstruction, in the solid green line, and the ground truth video, in the dashed gray line. The hotspot position angle and flux are calculated as described in Section \ref{subsec:simple_imaging}. The right panel shows the time-averaged hotspot flux in the red squares and the time-average of the residual of the hotspot position from ground truth in the blue circles for reconstructions with different $\lambda_{\rm OT}$ hyperparameter values. The vertical dashed red line denotes the reconstruction with maximum average hotspot flux, at $\lambda_{\rm OT} = 100$, and the vertical dashed blue line denotes the reconstruction with minimum average hotspot position angle residual, at $\lambda_{\rm OT} = 80$.}
    \label{fig:hs_properties}
\end{figure*}

Before performing full imaging on realistic data, we first demonstrate the power of the OT regularizer on data without gain corruptions. In this section, we use only the geometric \texttt{m-ring} and clockwise hotspot model.

To isolate the effect of the OT regularizer, we do not use the standard dynamic regularizers. We refer to solving Equation \ref{eq:dynamicRML} without the standard dynamic regularizers (i.e.,\ by  setting $\lambda_{\Delta t} = \lambda_{\Delta I} = 0$) via the L-BFGS algorithm as \textit{OT imaging}. As the initial video input to the optimization algorithm, we use a static video with each frame equal to the temporally averaged ground truth video, which is approximately the static \texttt{m-ring} model. Here, we vary $\lambda_{\rm OT}$ between $10^{-3}$ and $10^3$. It would also be possible to vary $\varepsilon$, the Sinkhorn entropy term weight, and $n_{\rm sh}$, the number of iterations for which the Sinkhorn algorithm is run at each evaluation of the OT regularizer. However, we find that these both have minor impacts on the resulting reconstructions, so we fix these to $\varepsilon = 10^{-4}$ and $n_{\rm sh} = 100$. Since the data is simple, we only perform one round of OT imaging for 50 iterations of the optimization algorithm. A round of OT imaging for one parameter set takes approximately two minutes on a single CPU core with 8 GB of memory. For comparison, an identical round of imaging with only the standard regularizers takes approximately 20 seconds on the same computational hardware. Our full, parallelized OT imaging survey over 22 values of $\lambda_{\rm OT}$ takes approximately 12 minutes on 16 CPU cores, each with 16 GB of memory.

We find that for this simple test, the OT regularizer performs well at resolving orbital motion. The fiducial reconstruction, with $\lambda_{\rm OT} = 80$, is shown in Figure \ref{fig:test_recon}. As shown in the left panel of Figure \ref{fig:hs_properties}, it closely tracks the correct position angle of the hotspot at most times, although the hotspot flux is typically well below the true value. This is expected in such a low number of optimization iterations, because we start with a static video. Despite the sparse 2017 $(u,v)$ coverage of {\sgra} and the time-dependent evolution of the asymmetry signal-to-noise ratio (SNR), the dynamic reconstruction, which uses the full observing time window rather than independent snapshots, achieves good overall accuracy after a single round of OT imaging, although the position angle is not well recovered at some epochs, such as near 12.6 and 13.5 UTC.

To demonstrate the effect of the OT regularizer hyperparameter value on the fidelity of the reconstructions, the right panel of Figure \ref{fig:hs_properties} shows the mean hotspot flux and mean residual position angle between the reconstructed and ground truth videos for different $\lambda_{\rm OT}$ values. We see that $\lambda_{\rm OT}$ values between about 60 and 100 perform the best at tracking the hotspot, with $\lambda_{\rm OT} = 100$ having the highest mean hotspot flux and $\lambda_{\rm OT} = 70$ having the lowest mean position angle residual. Lower $\lambda_{\rm OT}$ values are likely unable to strongly control for smoothness of the orbital motion, which explains the significantly larger mean position angle residual for $\lambda_{\rm OT} \leq 1$. On the other hand, higher $\lambda_{\rm OT}$ values cause the OT regularizer to dominate the data term, which imprints the dynamics of the orbiting hotspot, so that motion between frames is suppressed and the hotspot has a much lower mean flux. This also suggests that $\rho_{\rm DNX}$ is indeed an appropriate metric for selecting the fiducial parameter set when the ground truth video is known, as the fiducial reconstruction with $\lambda_{\rm OT} = 80$ has both a low mean position angle residual and a high mean hotspot flux.

In Figure \ref{fig:test_recon}, we see that, compared to the reconstructions with $\lambda_{\rm OT} = 10^{-3}$ and $10^3$, the hotspot in the fiducial reconstruction with $\lambda_{\rm OT} = 80$ has a greater flux, and its location is more correlated to the ground truth hotspot. The times selected in this figure are those with the highest SNR, where we would expect the reconstruction fidelity to be best, but the pattern holds true across the entire reconstruction window.

\begin{figure*}[htb]
    \centering
    \begin{minipage}[c]{\linewidth}
    \centering
        \includegraphics[width=0.98\linewidth]{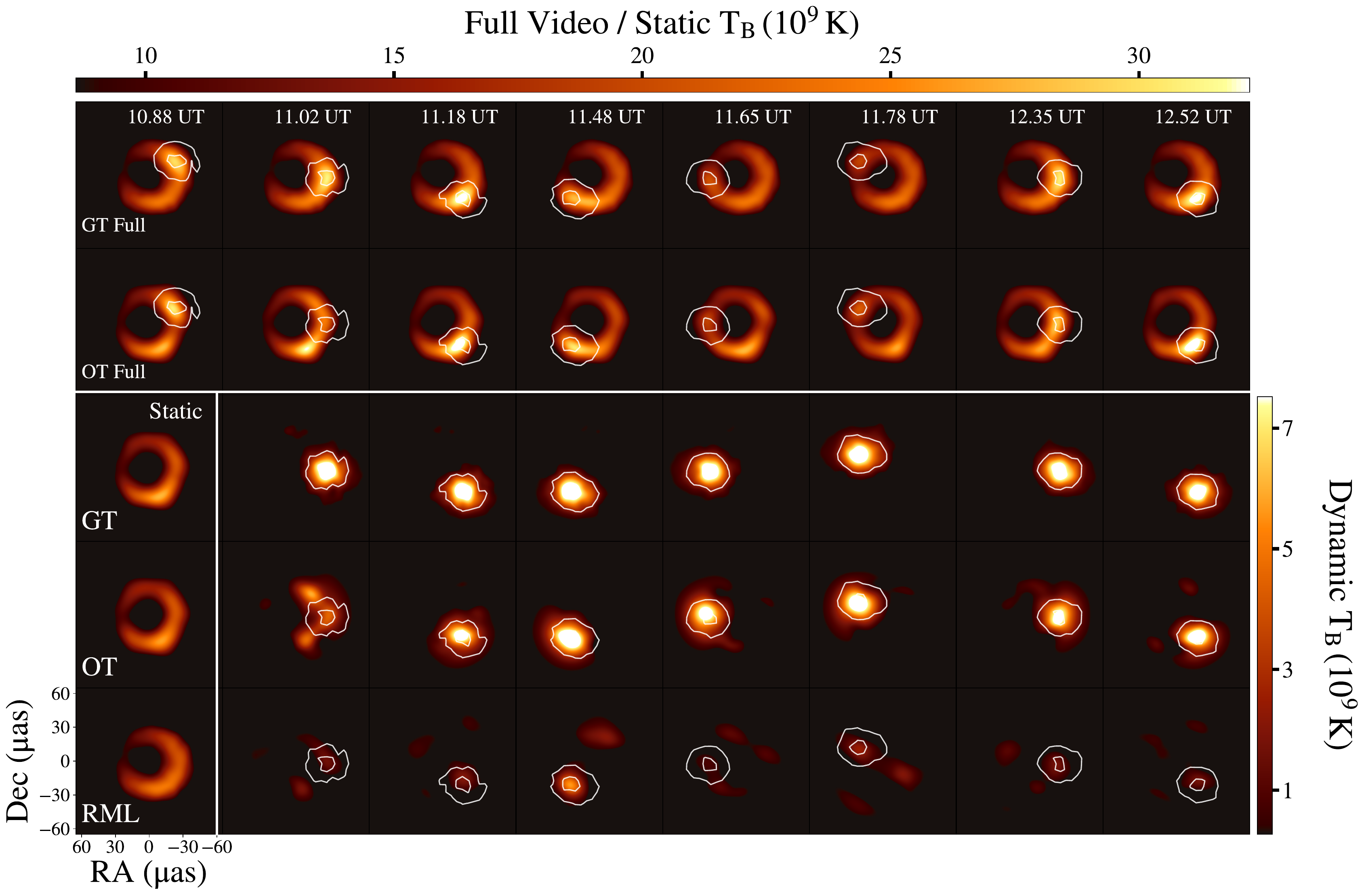} \\
        \includegraphics[width=0.75\linewidth]{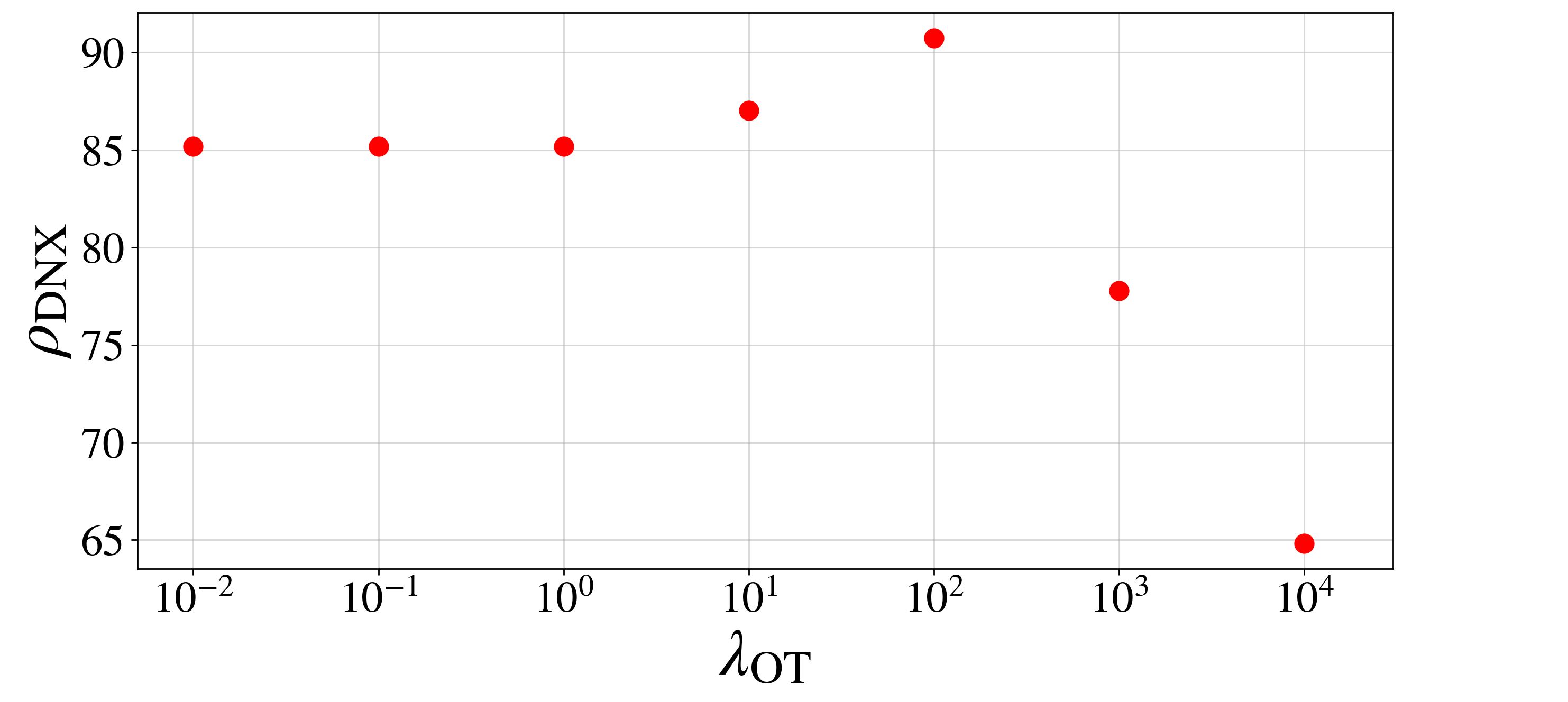} \\
        
    \end{minipage}
    \caption{Results from full imaging of the \texttt{m-ring} and clockwise hotspot geometric model. In the top panel, the first row shows the ground truth video. The second row shows the fiducial OT reconstruction, which is the reconstruction from OT imaging with maximum $\rho_{\rm DNX}$, corresponding to $\lambda_{\rm OT} = 100$. The bottom three rows show the static components (the median frame) in the first column and the dynamic components (residuals from the median frame) in the following columns of the ground truth video, the fiducial OT reconstruction, and the standard reconstruction, respectively. The standard reconstruction uses only the standard dynamic regularizers, and its temporal average is the initial video input for OT imaging. The non-static columns are spaced equally in terms of frame numbers, with four frames between each column, but not necessarily in time because there are time gaps in the data. The nominal time interval for each frame is two minutes (0.033 hours). The white contours denote the position of the ground truth hotspot, which is not perfectly circular because this synthetic data includes scattering effects. The bottom panel shows dynamic normalized cross-correlation ($\rho_{\rm DNX}$) scores of the OT reconstructions for different values of $\lambda_{\rm OT}$.}
    \label{fig:fullreconstruction}
\end{figure*}

\begin{figure*}[htb]
    \centering
    \begin{minipage}[c]{\linewidth}
    \centering
        \includegraphics[width=0.98\linewidth]{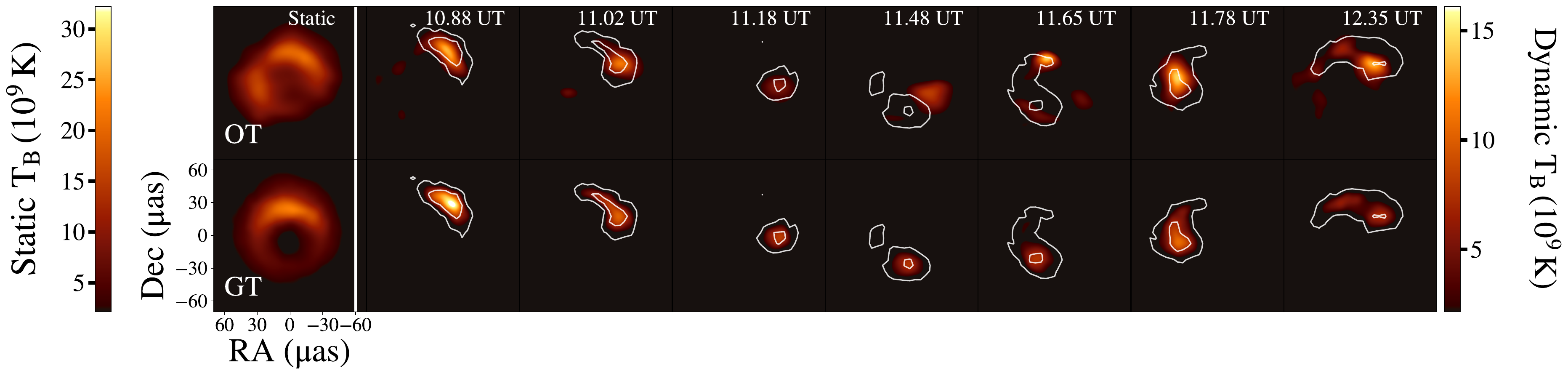}
    \end{minipage}
    \caption{Results from full imaging of the GRMHD and clockwise hotspot model. The first row shows the fiducial OT reconstruction and the second row shows the ground truth video, with the same format as the corresponding rows of Figure \ref{fig:fullreconstruction}. The columns are spaced equally in terms of frame numbers, with four frames between each column, but not necessarily in time because there are time gaps in the data. The nominal time interval for each frame is two minutes. The white contours denote the dynamic component of the ground truth video, which does not consist of only the hotspot due to the dynamic nature of the GRMHD model.}
    \label{fig:grmhd_reconstruction}
\end{figure*}

\subsection{Full Imaging Demonstration}
\label{subsec:full_imaging}

In this section, we perform full imaging on ``on-sky'' synthetic data, which includes diffractive and refractive scattering effects, complex amplitude and phase gain corruptions, and thermal noise. We use both the geometric and GRMHD models.

A full parameter survey on this data is unfeasible with the computational complexity of the OT regularizer, despite the relative efficiency of the Sinkhorn algorithm. Therefore, when performing our full imaging pipeline, we first run a larger parameter survey using only the standard dynamic regularizers, then subsequently run a smaller OT imaging parameter survey. We henceforth refer to solving Equation \ref{eq:dynamicRML} with only the standard dynamic regularizers (i.e.,\ by setting $\lambda_{\rm OT} = 0$) via the L-BFGS algorithm as \textit{standard imaging}. Parameters surveyed in standard imaging include the hyperparameters of the static imaging regularizers and the standard dynamic regularizers, a blurring kernel size, and the relative weighting between the data product terms in the objective function. For additional details on the standard imaging steps, see Section 3 of \citet{ehtim_official}.

Once the standard imaging parameter survey is complete, we evaluate the resulting reconstructed videos based on their $\rho_{\rm DNX}$ scores. We select the best-performing video, which we refer to as the \textit{standard reconstruction}, and the corresponding parameter set to undergo OT imaging. The initial video input to the OT imaging parameter survey is the temporally averaged standard reconstruction. We also test using the standard reconstruction itself, but find no significant difference in the resulting outputs. For this test, we vary the OT regularizer hyperparameter between $10^{-2}$ and $10^4$. Apart from the standard regularizer hyperparameters, which are again set to zero, all other imaging parameters are kept the same as in the standard reconstruction parameter set. We again fix $\varepsilon = 10^{-4}$, but we choose a more conservative number of Sinkhorn algorithm iterations of $n_{\rm sh} = 300$, which may increase the computational time but ensures that the Sinkhorn algorithm converges for each evaluation.

Taking the parameters as described in the paragraph above as input, we iteratively perform OT imaging, remove artifacts, and smooth temporally and spatially for a total of 10 rounds. Each of the 10 rounds of OT imaging includes 50 iterations of the L-BFGS optimization algorithm. We refer to the final output videos as the \textit{OT reconstructions}. The best OT reconstruction, as evaluated by $\rho_{\rm DNX}$, is referred to as the \textit{fiducial OT reconstruction}. The parameters corresponding to the fiducial OT reconstruction are the \textit{fiducial parameter set}. 

We find that the OT reconstructions for the \texttt{m-ring} and clockwise hotspot geometric model are highly correlated with the ground truth videos and far outperform the standard reconstruction at recovering orbital motion. In Figure \ref{fig:fullreconstruction}, we show the fiducial OT reconstruction in the second row and its static and dynamic components in the fourth row. The fiducial OT reconstruction, which corresponds to $\lambda_{\rm OT} = 100$, qualitatively looks similar to the ground truth video, shown in the first and third rows, with similar flux values as well. The times selected in the figure cover a range of SNR values, but with the exception of the frame at 11.02 UT, the dynamic component of the fiducial OT reconstruction indeed tracks the true hotspot location, indicating a clear ability to recover coherent motion.

We also see a marked improvement over the standard reconstruction, shown in the bottom row of Figure \ref{fig:fullreconstruction}. Although the static \texttt{m-ring} structure is present in the standard reconstruction, the recovered dynamic component is significantly degraded compared to that of the fiducial OT reconstruction. The standard reconstruction dynamic component flux is lower across all frames, and while the location of the dynamic component appears to have some correlation with the true hotspot location, it is unclear in most frames. The presence of many artifacts also make it so that, without the ground truth hotspot contours to guide the eye, the coherent motion is not clearly visible in the standard reconstruction.

Using the same method to extract the hotspot flux and position angle as in Section \ref{subsec:simple_imaging}, we again find that the fiducial OT reconstruction tracks the ground truth hotspot far better than the standard reconstruction. The flux of the ground truth hotspot is approximately stable around 0.097 Jy; in comparison, the mean hotspot flux in the fiducial OT reconstruction is 0.084 Jy and only 0.016 Jy in the standard reconstruction. Whereas the hotspot flux in the standard reconstruction never exceeds 0.4 times the ground truth hotspot flux, 84\% of frames in the fiducial OT reconstruction have hotspot fluxes greater than 0.8 times the ground truth hotspot flux. 

The hotspot position angle residual between the reconstructed and ground truth videos is also improved by the addition of the OT regularizer, with an average of 6.3$^\circ$ for the fiducial OT reconstruction and an average of 21.1$^\circ$ for the standard reconstruction. For 80\% of frames in the fiducial OT reconstruction and only 55\% of frames in the standard reconstruction, the hotspot position angle is within 10$^\circ$ of the true position. Thus, it is clear that OT regularization is essential to successfully reconstruct coherent motion to high fidelity in an RML framework.

As in the simple imaging test described in Section \ref{subsec:simple_imaging}, we find that this dataset also achieves the best performance at $\lambda_{\rm OT} \approx 100$. 
For significantly higher or lower values of this hyperparameter, the constraint on smooth motion becomes, respectively, too strong or too weak. Because the OT regularizer is computed on normalized frames, its absolute scale is independent of the total source flux. 
Moreover, Figure \ref{fig:otpwGRMHD} shows that it is also largely insensitive to the effective angular resolution of the array. 
The full imaging pipeline performs similarly well for both the GRMHD and hotspot models. 
Figure \ref{fig:grmhd_reconstruction} shows that the temporal evolution of the morphology in the fiducial OT reconstruction closely follows that of the ground truth video, and that the brightest region is robustly tracked in the reconstruction. 
However, some smaller-scale asymmetries appear displaced or blurred in the OT reconstruction, indicating that while the reconstruction successfully recovers the global structure and motion, limitations still remain in the fidelity of the small-scale structure.

Taken together, these results show that the dynamics can be detected in both the geometric and GRMHD models. 
In particular, the value $\lambda_{\rm OT}=100$ identified in this study provides good performance for the 2017 April 11 coverage. 
Furthermore, provided that the data quality is not substantially different from that of the present observations, this optimal value is expected to remain effective under similar observing conditions. 
Looking ahead, if future EHT observations benefit from improved $(u, v)$ coverage and higher SNR through the addition of new telescopes and enhanced observational capabilities, it will be worthwhile to repeat the parameter survey presented here using simulated data based on those updated conditions to investigate whether the detection of dynamics can be further improved. 
This represents an important direction for future work.

\section{Conclusion}
\label{sec:conclusion}

This work presents the first proposal to apply OT-based methods to radio interferometric imaging. We find that the OT regularizer in an RML framework performs remarkably well in our imaging tests, which consist of synthetic observations of a simple geometric model and a GRMHD model.
The power of the OT regularizer lies in the transport matrix of the OT calculation. By explicitly, but flexibly, modeling the flow of intensity between frames via this matrix, OT captures the underlying dynamics of the source rather than relying on rigid pixel-based correlation assumptions.
We have implemented this method as a new extension to the official EHT imaging software, $\texttt{eht-imaging}$ (\texttt{ehtim+OT}; \citealt{ehtim_official}), contributing to the advancement of the dynamic analysis pipeline within the EHT Collaboration.
The theoretical formulation and numerical validation presented in this
work form the foundation for the implementation and verification tests
described in another paper \citep{ehtim_official}.
The OT regularizer proposed in this study has been integrated into the official EHT imaging pipeline, providing a fully operational framework for dynamic reconstructions of {\sgra} based on EHT observations.
Furthermore, the OT regularizer is planned to be implemented in the next-generation Julia-based analysis framework of the EHT (\texttt{EHTJulia}; \url{https://github.com/EHTJulia}), which is expected to enable a faster and more flexible dynamic imaging environment.

In the companion \texttt{ehtim+OT} paper, a series of benchmark tests using a variety of dynamic models demonstrate that the OT regularizer achieves high-fidelity reconstructions while maintaining temporal consistency across frames.
The entire pipeline has been designed with reproducibility and verifiability in mind, ensuring consistency across the EHT Collaboration and allowing for straightforward extension to future dynamic imaging analyses with the next-generation EHT \citep{ngeht_paper} and the Black Hole Explorer \citep{johnson_2024_bhex}.
Together, this work and the \texttt{ehtim+OT} pipeline built upon it provide a unified framework for probing the time-variable phenomena near black holes with unprecedented accuracy, combining theoretical methodology and software implementation.

There are several promising directions for future extensions of this approach.
One is to determine the optimal number of frames, or equivalently the time resolution, for dynamic reconstruction.
A key difficulty in VLBI imaging arises from the sparse $(u, v)$ coverage, and dynamic imaging exacerbates this issue due to the requirement of dividing the data into multiple frames.
By incorporating the OT regularizer, it may be possible to reduce the number of frames, and thus increase the time intervals and $(u,v)$ coverage of each individual frame, since the OT regularizer can account for the underlying dynamics.

Another possible extension involves redefining the OT distance.
In the current implementation, we use the squared Euclidean distance as the cost matrix for the OT problem; however, the distance function can be modified to favor circular motion. For example, a distance function that penalizes radial motion separately from orbital motion could be applicable to these data. 
Also, for our definition of the Sinkhorn distance, $W_\varepsilon(\bm{I}, \bm{I})$ is not necessarily equal to zero for a general $\bm{I}$; to ensure that the distance is a proper metric, we could use the Sinkhorn divergence \citep{ramdas_wasserstein_2017}, which introduces debiasing terms to satisfy the coincidence property. The Sinkhorn divergence may be helpful in cases where adjacent frames are very similar; however, a primary motivation for using the OT regularizer is that it remains a strong temporal regularizer even when the separation between frames is large relative to the source variability time scale. Furthermore, calculating the Sinkhorn divergence requires at least double the number of calls to the Sinkhorn algorithm, making it less efficient.
Several other formulations of the OT distance, such as the relaxed OT distance for non-normalized images \citep{frogner_learning_2015, chizat_2017_sclaing} or the sharp Sinkhorn distance \citep{luise_differential_2018}, exist as well, and although we do not find any significant change in the results of our imaging test when using these variations, further exploration of these possibilities on other data is left for future work.

\begin{acknowledgements}
This work was supported by JSPS KAKENHI Grant Numbers JP23K20035 and JP24H00004 (to S.I.).
\end{acknowledgements}

\end{document}